\documentclass[12pt, referee]{aastex}
\usepackage[dvips]{color}

\newcommand{\bl}{\textcolor[rgb]{0,0,0}}
\newcommand{\bu}{\textcolor[rgb]{0,0,0}}
\newcommand{\ma}{\textcolor[rgb]{0.0,0,0.0}}

\def\gtsima{$\;\buildrel > \over \sim \;$}
\def\simgt{\lower.5ex \hbox{\gtsima}}
\def\ltsima{$\;\buildrel < \over \sim \;$}
\def\simlt{\lower.5ex \hbox{\ltsima}}

\begin{document}

\title{SOFIA/EXES observations of warm H$_2$ at high spectral resolution: witnessing para-to-ortho conversion behind a molecular shock wave in HH7}

\author{David A. Neufeld\altaffilmark{1}, Curtis DeWitt\altaffilmark{2}, Pierre Lesaffre\altaffilmark{3,4}, Sylvie Cabrit\altaffilmark{4}, Antoine Gusdorf\altaffilmark{3,4}, \\
\bu{Le Ngoc Tram\altaffilmark{2,5}}, and Matthew Richter\altaffilmark{6}}

\altaffiltext{1}{Department of Physics \& Astronomy, Johns Hopkins Univ., Baltimore, MD 21218, USA}
\altaffiltext{2}{SOFIA Science Center, NASA Ames Research Center, Moffett Field, CA 94035, USA}
\altaffiltext{3}{\bu{Laboratoire de Physique de l`\'Ecole normale sup\'erieure, ENS, Universit\'e PSL, CNRS, Sorbonne Universit\'e, Universit\'e de Paris, France}}
\altaffiltext{4}{Observatoire de Paris, PSL University, Sorbonne Universit\'e, LERMA, 75014 Paris, France}
\altaffiltext{5}{University of Science and Technology of Hanoi (USTH), Hanoi, Vietnam}
\altaffiltext{6}{Universiy of California, Davis, CA 95616, USA}

\begin{abstract}

\bl{Spectrally-resolved observations of three pure rotational lines of H$_2$, conducted with the EXES instrument on SOFIA toward the classic bow shock HH7, reveal systematic velocity shifts between the S(5) line of ortho-H$_2$ and the two para-H$_2$ lines [S(4) and S(6)] lying immediately above and below it on the rotational ladder. These shifts, reported 
here for the first time, imply that we are witnessing the conversion of para-H$_2$ to ortho-H$_2$ within a shock wave 
driven by an outflow from a young stellar object.  The observations are in good agreement with the predictions of
models for non-dissociative, C-type molecular shocks.  They provide a { clear demonstration} of the chemical changes wrought
by interstellar shock waves, in this case the conversion of para-H$_2$ to ortho-H$_2$ in reactive collisions 
with atomic hydrogen, and { provide among the most compelling evidence yet obtained for C-type shocks in which 
the flow velocity changes continuously.}}

\end{abstract}


\section{Introduction}

Shock waves are a widespread phenomenon in the interstellar medium; they may be driven by supersonic motions associated with protostellar outflows, supernova explosions, and cloud-cloud collisions.  In addition to heating and compressing the medium through which they propagate, shock waves can alter the chemical state of the interstellar gas.  Even in slower shocks that are non-dissociative, the passage of a shock wave through a molecular cloud can cause significant changes in its composition, either as a result of grain mantle sputtering, or because of large enhancements in the rates of chemical reactions that are endothermic or possess an activation energy barrier  \bl{(Godard et al.\ 2019, and references therein.)}  One theorized example of such a change is the conversion of para-H$_2$ to ortho-H$_2$ behind a \bl{molecular} shock wave (e.g. Timmermann 1998; Wilgenbus et al.\ 2000). 

Because their interconversion by means of radiative processes or in non-reactive collisions is negligibly slow, ortho- and para-H$_2$ may be regarded as two distinct chemical species.  On long timescales (\bl{$\sim 0.1$ Myr in dark clouds; e.g.\ Harju et al.\ 2017}), reactions with H$_3^+$, H$^+$, and H -- involving the breaking of chemical bonds -- can lead to ortho-para conversion and thereby establish an equilibrium ortho-to-para ratio (OPR) that is determined by the gas temperature.  The OPR in local thermodynamic equilibrium (LTE) is temperature-dependent, because of the intimate connection between the rotational state of the molecule and the spin state of the nuclei.  If the spin wavefunction is anti-symmetric (para-H$_2$ \bl{with total nuclear spin 0}), then the rotational wavefunction must be symmetric and the rotational quantum number, $J$, must be even.  Conversely, rotational states of ortho-H$_2$ (\bl{total nuclear spin 1}) all have odd $J$.  In the limit of high temperature, multiple states of both ortho- and para-H$_2$ are populated, and the OPR approaches 3, the ratio of the nuclear spin degeneracies.  In the opposite limit of low temperature \bl{(i.e. for $T \ll E_1/k = 170\,\rm K$, where $E_1$ is the energy of the $J=1$ rotational state)}, only the lowest $J=0$ state is significantly populated and the OPR approaches zero.  In LTE, OPRs of 1, 2, and 2.9 are achieved at gas kinetic temperatures of 78, 118, and 216~K, respectively.

Observations of the H$_2$ OPR within shock-heated molecular gas became possible with the {\it Infrared Space Observatory} ({\it ISO}).   Through observations of the H$_2$ S(1) -- S(5) pure rotational lines toward shocked gas in HH54, {\it ISO} revealed an OPR $\sim 1.2$ in gas of kinetic temperature $\sim 650$~K (Neufeld et al.\ 1998).  This measurement of a non-equilibrium OPR in warm gas for which the OPR would be 3 in LTE implied that para-to-ortho conversion, if it occurred at all, did not proceed to completion within the time period for which the emitting gas was warm.  Thus, the warm H$_2$ observed by {\it ISO} retained an OPR that was a relic of an earlier phase in which the gas had equilibrated at a lower temperature.  \bl{Non-equilibrium OPRs were also measured toward HH2 by {\it ISO} (Lefloch et al.\ 2003), and were in good agreement with the predictions of Wilgenbus et al.\ (2000).} 

Later observations with the IRS instrument on {\it Spitzer} allowed the H$_2$ S(0) -- S(7) line intensities to be measured toward multiple sources with better sensitivity than {\it ISO} (e.g.\ Neufeld et al.\ 2006, 2007, 2009).  Such measurements are conveniently represented in
rotational diagrams, obtained by plotting ${\rm log}(N_J/g_J)$ versus $E_J$, where $N_J$ is the column density in rotational state $J$, $g_J$ is the degeneracy ($2J+1$ for even $J$ and $3(2J+1)$ for odd $J$) and $E_J$ is the rotational energy. The rotational diagrams obtained with {\it Spitzer} typically exhibited three key behaviors.  First, positive curvatures in the rotational diagrams indicated the presence of multiple gas temperatures, suggesting that
shocks with a range of velocities were present within the beam.  Second, zigzag patterns in the rotational diagrams -- with ${\rm log}(N_J/g_J)$ systematically higher for states of even $J$ -- implied an OPR below the value expected in LTE.  And third, the departure from OPR (i.e. the degree of the zigzag) was typically greater for the lower rotational states than for the higher ones.  This third feature of the rotational diagrams suggested indirectly that para-to-ortho conversion was taking place within the shocks that had been observed, and that the faster shocks responsible for the higher-$J$ line emission were more efficient in converting para-H$_2$ to ortho-H$_2$.  The greater efficiency of faster shocks could be readily understood in the context of shock models, which indicated that para-to-ortho conversion is dominated by reactive collisions with atomic H that possess an activation energy barrier corresponding to a temperature of $\sim 4000$~K (Schofield 1967; Lique 2015). In this picture, faster shocks -- which can be partially dissociative -- produce more atomic hydrogen which can react more quickly with H$_2$ at the higher postshock temperatures.

Despite the success of shock models in explaining the H$_2$ rotational diagrams obtained with {\it Spitzer}, direct evidence for para-to-ortho conversion has been elusive.  Even in the nearest sources, para-to-ortho conversion occurs on a length scale that is too small\footnote{\bl{The length scale is $\sim 10^{-3} (10^4 \, {\rm cm}^{-3} / n_0)\,\rm pc,$ where $n_0$ is the preshock density of H nuclei.  For HH7, which is among the most favorable sources ($n_0 = 10^4 \, \rm cm^{-3}$, distance $\sim 250$~pc), this corresponds to an angular scale of $\simlt 1^{\prime\prime}$.}} to be resolved with {\it Spitzer} (although future JWST observations may provide the necessary angular resolution).  But in addition to spatial offsets between the ortho- and para-H$_2$ emissions from shocks in which para-to-ortho conversion is occurring, shock models also predict small velocity shifts of a few km~s$^{-1}$.   { These shifts occur because the gas is being decelerated while the ortho-to-para ratio is increasing (Wilgenbus et al.\ 2000); thus, the ortho-H$_2$ emissions emerge preferentially from a region where the gas has been more significantly decelerated with respect to the preshock material.}  While the spectral resolution of {\it Spitzer}/IRS ($\lambda/\Delta \lambda \sim 60$ at the wavelengths of the S(3) -- S(5) lines) was far too poor to permit the detection of such small velocity shifts, ground-based observations of the para-H$_2$ S(4) line toward shocked gas in HH54 have revealed an intriguing shift relative to the ortho-H$_2$ S(9) pure rotational line and the $v = 1-0$~ S(1) line (Santangelo et al.\ 2014).  However, because the latter two lines are both of much higher excitation than the S(4) line, the velocity difference might be related to \bl{gradients in} excitation rather than to spin-symmetry; \bl{moreover, the slit position used for the H$_2$ S(4) observation was different from that used for the ortho-H$_2$ observations, making a direct comparison difficult.}

With the advent of the EXES spectrometer (Richter et al.\ 2018) on SOFIA, it is now possible to observe the S(4), S(5), S(6), and S(7) pure rotational lines of H$_2$ \bl{in the same observational configuration and} at a spectral resolution sufficient to resolve the expected ortho-para velocity shifts.  In this paper, we discuss the first unequivocal detection of such shifts, obtained toward HH7, a classic bow shock \bl{located in 
the NGC1333 cloud in the Perseus complex, where it is} driven by a jet from the young stellar object (YSO) SVS 13
\bl{(Bachiller et al.\ 1990)}.  The observations and data reduction are described in Section 2, and the results presented in Section 3.  A discussion follows in Section 4.

\section{Observations and data reduction}

We observed the S(4), S(5), S(6), and S(7) pure rotational transitions of H$_2$ toward two positions in HH7.
The observations were carried out using the EXES instrument in High-Medium mode, with a slit of width $1.9^{\prime\prime}$ centered on $[\alpha,\delta]=\rm 3^h29^m08\fs 46$,+31$^0$15$^\prime 29\farcs 2$ \bu{(J2000)} and $\rm 3^h29^m08 \fs 38$,+31$^0$15$^\prime 26\farcs 3$ \bu{(J2000)}.   
These positions, which we denote P1 and P2 respectively, are located 
\bu{close to the peak of HH7A (i.e.\ near the apex of the bow shock, as shown in Figure 1).  The offsets relative to the putative driving source, SVS 13A/VLA4B, were $(60\farcs 3, -34\farcs 8)$ and $(59\farcs 3, -37\farcs 7)$ respectively.}
\bl{The SOFIA image size at these wavelengths is typically $\sim 3.8^{\prime\prime}$ (50$\%$ encircled power; Temi et al.\ 2018)}. 
\begin{figure}
\includegraphics[scale=0.7,angle=0]{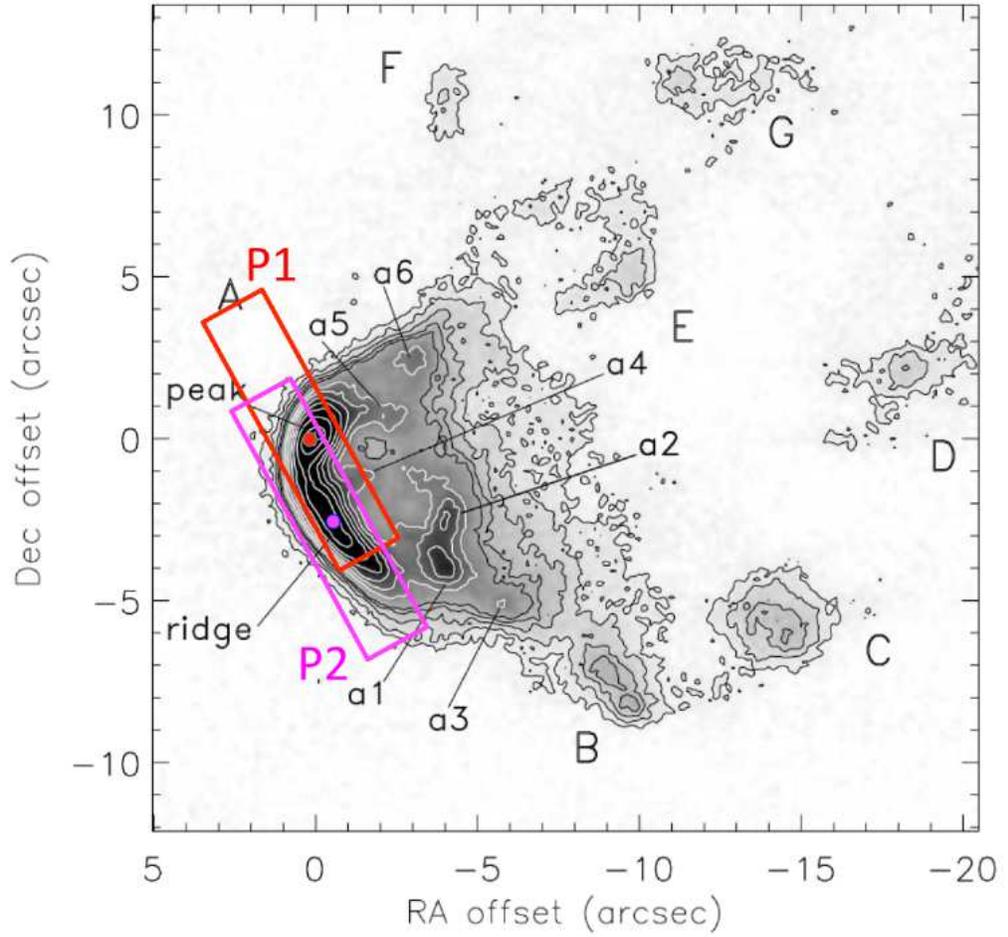}
\caption{Slit locations overlaid on the H$_2$ v$= 1 - 0$~\bu{S(1)} map of Khanzadyan et al.\ (2003). The slit position angle is shown
at the average value during the observations.}
\end{figure}

This configuration of EXES provides a spectral resolving power $\lambda/\Delta \lambda$ of 86,000, corresponding to a Doppler velocity of 3.5~km~s$^{-1}$.  In Table 1, we list for each spectral line the rest wavelength, slit length, date(s) of observation, integration time, and slit rotation angle (degrees East of North).  The latter varied during the observations, so a range is given.   
Figure 1 shows the average slit locations overlaid on the H$_2$ v$= 1 - 0$~\bu{S(1)} map of Khanzadyan et al.\ (2003). 
The telescope was nodded  after every \bu{52 -- 68}~s of integration 
to a reference position devoid of H$_2$ emission, enabling the subtraction of telluric emission features.

\begin{table}
\caption{\ma{Observational details}}
\vskip 0.1 true in
\begin{tabular}{lcccc}
\hline
Line 						& S(4) 		& S(5) 		& S(6) 		& S(7)		 	\\
Rest wavelength ($\mu$m)		& 8.02504108	& 6.90950858	& 6.10856384	& 5.51118327		\\
Slit length (arcsec)			& 12.4		& 9.8		& 8.4		& 7.6			\\
Dates of observations (P1)$^a$	& Oct 19, 20   & Oct 19, 20   & Oct 20  	& Oct 19 			\\
Dates of observations (P2)$^a$	& Oct 24 		& Oct 24, 25   & Oct 24 		& Oct 25			\\
Integration time$^b$ (s) (P1)		& 1792		& 2150		& 1800		& 1250			\\	
Integration time$^b$ (s) (P2)		& 1024		& 2316		& 1280		& 640			\\	
Slit position angle$^c$ (P1) 		& 14.6 -- 25.3 & 22.4 -- 47.8 & 35.0 -- 47.1 & 29.2 -- 35.3 	\\
Slit position angle$^c$ (P2) 		& 30.3 -- 36.7 & 36.6 -- 46.7 & 12.6 -- 27.4 & 15.9 -- 17.3 	\\
Line centroid$^{d,e}$ (P1)  		& $5.81 \pm 0.18$ & $2.19 \pm 0.22$ & $5.26 \pm 0.28$ & $1.74 \pm 0.57$  \\
Line centroid$^{d,e}$ (P2)  		& $7.39 \pm 0.18$ & $3.51 \pm 0.25$ & $6.35 \pm 0.43$ & $2.50 \pm 0.72$  \\
Line width $^{d,f}$ (P1)			& $10.1 \pm 0.4$ & $14.4 \pm 0.6$ & $11.6 \pm 0.7$ & $19.3 \pm 1.5$  \\
Line width $^{d,f}$ (P2)			& $8.9 \pm 0.4$ & $14.3 \pm 0.6$ & $13.0 \pm 1.1$ & $19.1 \pm 1.9$  \\ 
Integrated intensity$^{d,g}$ (P1) 	& $4.30 \pm 0.19$ & $4.19 \pm 0.17$ & $1.84 \pm 0.11$ & \bu{$3.63 \pm 0.31$} \\
Integrated intensity$^{d,g}$ (P2) 	& $3.51 \pm 0.17$ & $3.57 \pm 0.16$ & $1.99 \pm 0.17$ & \bu{$4.52 \pm 0.31$} \\

\end{tabular}
\tablenotetext{a}{All 2018, UT}
\tablenotetext{b}{Detector integration times, on source}
\tablenotetext{c}{Degrees East of North}
\tablenotetext{d}{Gaussian fit to average spectrum over 8.4$^{\prime\prime}$ extraction region (6$^{\prime\prime}$ for S(7))}
\tablenotetext{e}{$\rm km\,s^{-1}$ with respect to the Local Standard of Rest}
\tablenotetext{f}{$\rm km\,s^{-1}$ full width at half maximum}
\tablenotetext{g}{in units of $10^{-4}\rm \, erg\, cm^{-2}\,s^{-1} sr^{-1}$}
\end{table}

The data were reduced using the Redux pipeline (Clarke et al. 2015) with the fspextool software package -- a modification of the Spextool package (Cushing et al. 2004) -- which performs wavelength calibration, and a custom python script for aperture extraction.  \bl{The absolute flux calibration is
accurate to} $\sim \rm 25\%$ and \bl{the relative flux calibration to} $\sim \rm 12.5\%$.
\bu{The wavelength calibration was obtained from observations of multiple atmospheric lines of water and 
methane within each bandpass: we conservatively estimate its accuracy as $0.3\, \rm km \, s^{-1}$.} 
The atmospheric transmission is expected to exceed \bu{90}$\%$ for all observed transitions.

\section{Results}

The observations led to clear detections of all four target lines toward both positions.  In the bottom two panels of 
Figure 2, we show the resultant spectra, after subtraction of a 
linear baseline, each normalized with respect to the peak intensity.  Here, the S(4) -- S(6) lines were extracted over a region of length $8.4^{\prime\prime}$, 
the largest extraction region common to all three lines.  The extraction region for the S(7) line had a length of 6$^{\prime\prime}$. 
\bl{ For the S(4) -- S(6) lines, we also present spectral extractions for three equal subregions of length $2.8^{\prime\prime}$
along the slit, which we designate P1A/P2A, P1B/P2B, and P1C/P2C from southwest to northeast; these are shown in the top six
panels of Figure 2.}
\begin{figure}
\includegraphics[scale=0.7,angle=0]{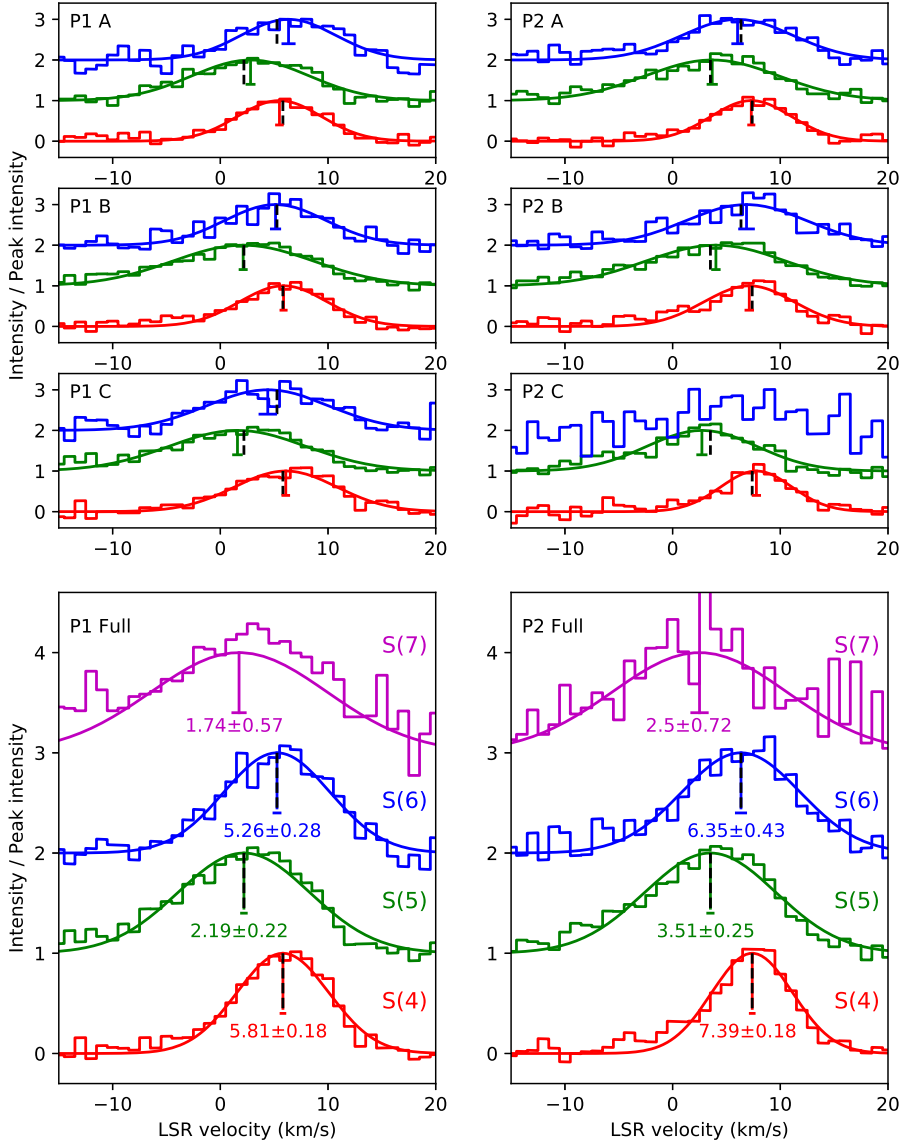}
\caption{$\rm H_2$ pure rotational spectra obtained with SOFIA/EXES toward HH7. 
The slit positions, extraction regions, and slit position angles are given in Table 1 or the text.  
The solid curves show Gaussian fits to the data plotted in the histogram.  { Vertical colored lines indicate 
the velocity centroids for each fit, with horizontal bars at the bottom indicating the standard errors 
(i.e. 68 \% confidence limits); dashed black vertical lines show for comparison the centroids obtained 
for the full slit extractions in bottom panels.}}
\end{figure}

Solid lines show the best Gaussian fits to the observed spectra; the velocity centroids, line widths (FWHM) and integrated
line fluxes for those fits are presented in Table 1, along with the standard errors on each parameter.  Both the fits and the observed data are normalized relative to the peaks of the Gaussian fits.  Vertical colored lines in Figure 2 indicate the velocity centroids for the Gaussian fits, \bl{with 
horizontal bars at the bottom indicating the standard errors (i.e. 68\% confidence limits).} 
As is immediately evident from even a cursory inspection of the bottom panels in Figure 2, both positions P1 and P2 
exhibit statistically-significant shifts between the S(5) ortho-H$_2$ line and two para-H$_2$ lines, S(4) and S(6).   
Because, the latter two lines "sandwich" the former in excitation level, $E_J$, the shift must be associated with 
the different spin symmetries rather than the excitation.  The measured magnitude of the ortho-para shift is 
\bl{ $3.62 \pm 0.28 $~km~s$^{-1}$ and $3.88 \pm 0.31 $~km~s$^{-1}$}, respectively, at positions P1 and P2. 
The errors given here are 1~$\sigma$ statistical uncertainties; the systematic uncertainties for each line position 
are \bu{less than} 0.3~km~s$^{-1}$.

The upper panels show that statistically significant velocity shifts between the S(4) 
and S(5) lines are detected within every subregion as well.  Here, the dashed black vertical lines show the velocity 
centroids obtained for the full $8.4^{\prime\prime}$ spectral extractions.  { The location of
these lines -- relative to the horizontal bars that show the standard errors on the velocity centroids for the subregions
-- indicate} that there is no statistically-significant 
evidence for a velocity gradient along the slit at either position P1 or P2; thus, { 
the observed ortho-para velocity shifts cannot be attributed to a chance combination of gradients in velocity 
and ambient (preshock) OPR along the slit.}

Unfortunately, at the time of the observations, the S(7) line was Doppler-shifted close to an order edge where 
the sensitivity is significantly reduced; here, the signal-to-noise ratio is insufficient to constrain the centroid 
velocity nearly as well as was possible for S(4), S(5) and S(6).\footnote{Owing to the US government shutdown, 
follow-up observations of H$_2$ S(7) -- scheduled for late January 2019 when the Doppler shift was considerably 
more favorable -- could not be performed.}  For the S(7) line, the shorter slit length prevented us from using the same extraction region as for the other lines.

\section{Discussion}
 
\subsection{Variation of the OPR behind the shock}

Our observations of HH7 provide the first definitive evidence for the conversion of para-H$_2$ to ortho-H$_2$ behind shock wave propagating in molecular gas.  Given a triad of observed line intensities for the H$_2$ S(4), S(5), and S(6) lines, we may determine uniquely the excitation temperature, $T_{86}$, implied by the S(6)/S(4) line ratio (which is proportional to the population ratio in $J=8$ and $J=6$), and the ortho-to-para, $\rm OPR_{678}$, that is needed to account for the S(5) line strength ($J=7$ level population) based on a linear interpolation of the rotational diagram between $J = 6$ and $J = 8$.  The results for both parameters are shown in Figure 3 (crosses) in velocity bins of width 1~km~s$^{-1}$.  The results are consistent with a steady increase of the OPR with increasing blueshift { (see the top panel of Fig.\ 4 below)}; this is the expected sense of the variation if the shock is propagating towards us into ambient molecular gas.  
 
\subsection{Comparison with shock models}

We have compared the observational results presented here with predictions from the Paris-Durham shock model (\bl{Lesaffre et al. 2013; Flower \& Pineau des For\^ets 2015; Godard et al.\ 2019}).  The upper panel of Figure 4 shows the gas kinetic temperature, the flow velocities for the ionized and neutral species, and the OPR predicted for a plane-parallel, steady-state C-type shock\footnote{Models for J-type (i.e.\ ``Jump"-type) shocks are unsuccessful in accounting for these data.  Because the gas is rapidly decelerated at the shock front within a J-type shock, the predicted line profiles are very narrow; hence, ortho-para velocity shifts are not expected from a single J-shock.   While a collection of multiple J-type shocks of varying \bu{shock} velocity might yield an ortho-para \bu{velocity} shift in their combined emission, this would be accompanied by a shift between the S(4) and S(6) lines 
because the postshock temperature (which determines the S(6)/S(4) line ratio) and degree of para-to-ortho conversion both increase with shock velocity.  \bu{These shock-velocity-dependent effects for J-shocks were considered quantitatively by Wilgenbus et al. (2000; their Figure 5b). Their analysis showed that an increase in the OPR from 0.1 to $ > 2$ -- resulting from an increase in the shock velocity -- should go hand in hand with a sharp increase in H$_2$ rotational temperature, leading to temperatures greater than 1500 K at shock velocities where the OPR $ > 2$.  No such increase in the rotational temperature with OPR was observed (Figure 3).}}
of preshock H nucleus density $n_0=10^4\,\rm cm^{-3}$, velocity $v_s=$ 25~km~s$^{-1}$, perpendicular preshock magnetic field $B_0= 130\,\rm \mu G$, and initial ortho-to-para ratio OPR$_0=0.01$.
Here, the shock velocity and preshock magnetic field 
\bl{have been adjusted to provide the optimal fit to the observed S(6)/S(4) line ratio, and the preshock density 
lies within the range of values derived in previous studies (Smith et al.\ 2003; Molinari et al.\ 2000; Yuan et al 2012).  The initial OPR is the value that would be expected if the preshock gas had achieved equilibrium at a temperature of 25~K;
significantly larger values for OPR$_0$ worsen the agreement with the measurements of OPR$_{678}$ plotted in Figure 3.  The shock was assumed to be irradiated by an ultraviolet radiation field with an intensity equal to the mean interstellar value.}

\begin{figure}
\includegraphics[scale=0.7,angle=0]{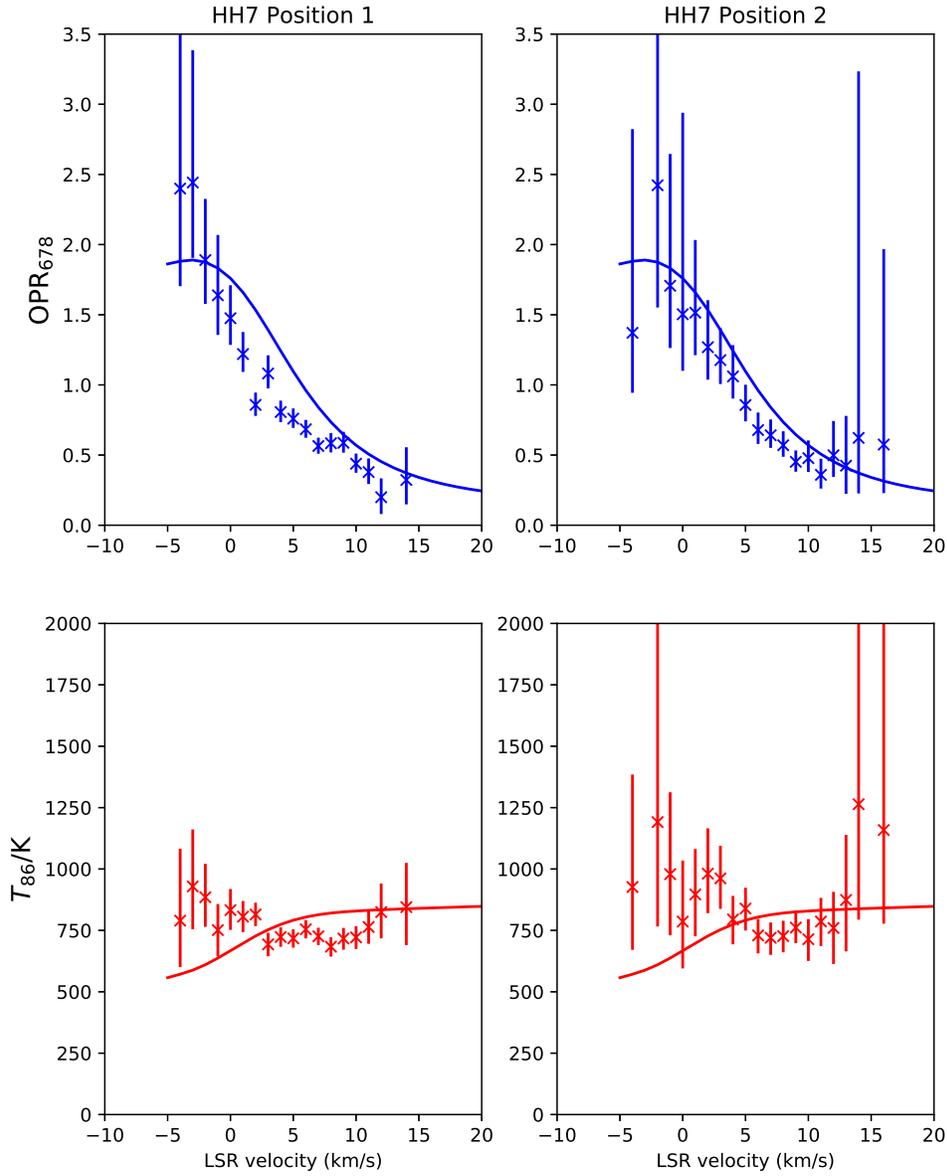}
\caption{Crosses with 68$\%$ confidence limits: variation of OPR$_{678}$ (upper panels) and $T_{86}$ (lower panels) with Doppler velocity.   Solid curves: predictions for the simple shock model described in the text. }
\end{figure}
\begin{figure}
\includegraphics[scale=0.75,angle=0]{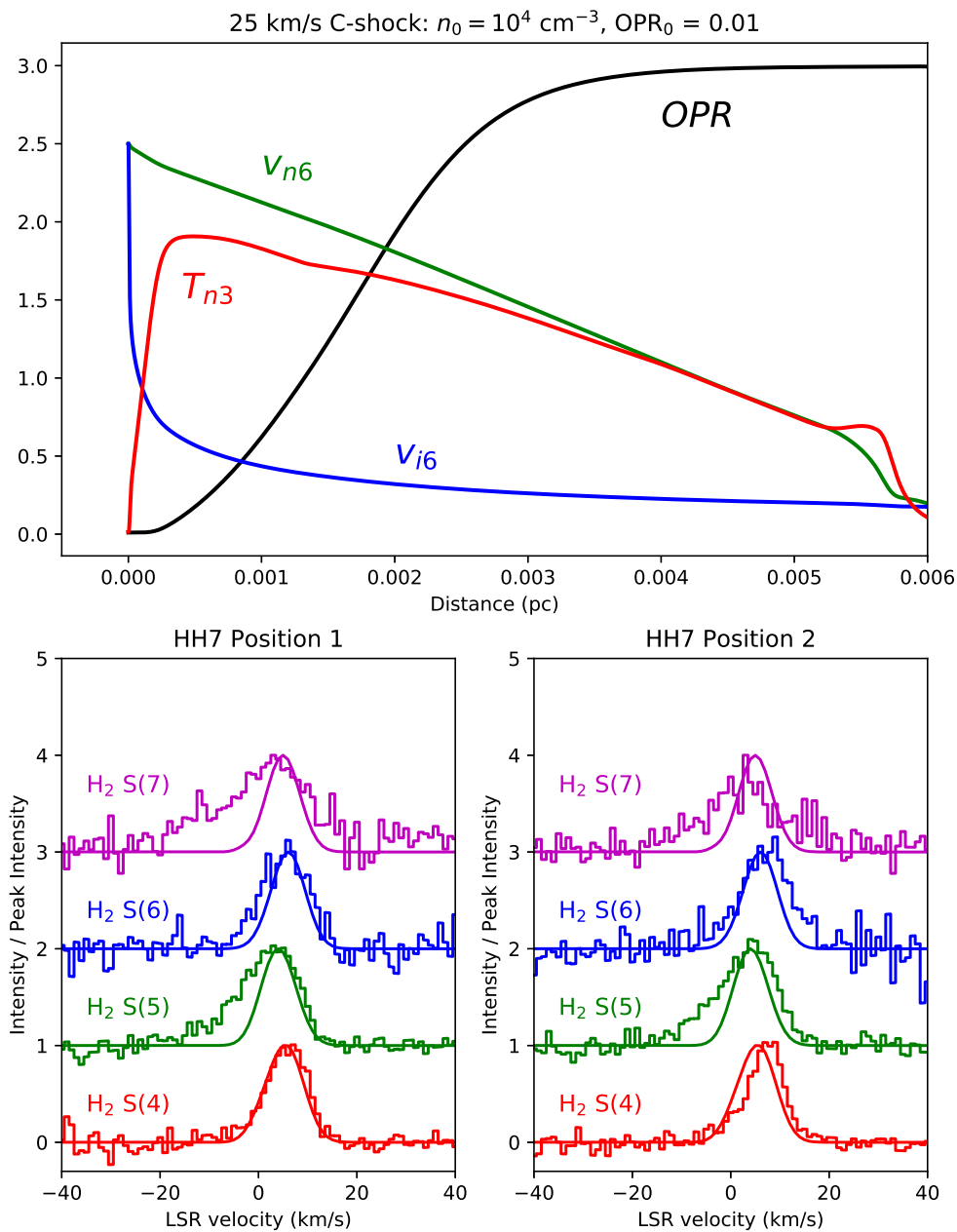}
\caption{Top panel: neutral gas temperature ($T_{3} = T/[10^3$~K]), velocities for the neutral and ionized fluids ($v_{n6,i6} = v_{n,i} / [10^6 \rm \, cm \,s^{-1}]$), and H$_2$ OPR for
the simple C-type shock model described in the text.  Bottom panels: predicted H$_2$ line profiles for the
simple model described in the text, overlaid on the 
observed spectra for 
positions P1 and P2.}
\end{figure}

In the middle and lower panels of Figure 4, we show the predicted line profiles, after convolution with the 
instrumental profile { (which is only of minor importance)}.
These were obtained using the methods described by Tram et al. (2018), and include the effects of gradients in velocity 
and OPR within the shock (upper panel) as well as thermal broadening.
Here, we adopted a LSR velocity of $8.4\, \rm km\,s^{-1}$ for the ambient material \bu{(Lefloch et al.\ 1998;
Lef\`evre et al.\ 2017)} and assumed the 
flow was inclined towards us at an angle of \bl{60$^0$} to the line-of-sight, 
consistent with the inclination angle of 30 - 60$^0$ estimated previously (Davis et al.\ 2000, and references therein).  { For this inclination angle, 
the ortho-para velocity shift in the predicted line profiles is $1.5\,\rm km\,s^{-1}$ (shift between the centroid of a Gaussian fit to the S(5) line and the mean of the centroids for S(4) and S(6)); and thermal broadening and the velocity gradient make comparable contributions to the overall line width.}

Given the simplicity of this model, in which a single plane-parallel shock is responsible for the observed emission, 
the fit to the data is reasonable, particularly at position P1.  Clearly, the S(5) -- S(7) lines
tend to show a blue excess relative to the predicted profile.  Most likely, this discrepancy reflects a  shortcoming of the simple model that we have adopted;
if, as seems probable, the slit contains a mixture of shocks of varying speed, then the faster shocks could create a blue excess (characterized by a larger temperature and OPR).  { These blue excesses, which are most pronounced for the S(5) and S(7) lines, account for part of the ortho-para shifts that are apparent in Figure 2. Nevertheless, an examination 
of the right half of each observed line profile in Position P1 clearly indicates that the ortho-para 
shifts in the predicted line profiles are needed to fit the data.}
A more complete analysis, which we defer to a future study, will take account of the three-dimensional structure of the bow shock in HH7, using methods similar those introduced by Tram et al.\ (2018), { and will attempt to model the different
observed line profiles at the two positions.} 

To enable a comparison with the results presented in Figure 3 for OPR$_{678}$ and $T_{86}$, we have treated the predicted line profiles in Figure 4 in exactly the same way as we did the observed line profiles, i.e.\
we computed the line ratios in $1\,\rm km\,s^{-1}$ velocity bins and derived OPR$_{678}$ and $T_{86}$ from those ratios.  The result is shown by the solid curves \bl{in Figure 3}.  The observed data points do not extend into the region of the blue excess ($v_{\rm LSR} \le -5 \, \rm km\,s^{-1}$), because the S(4) line intensity there is too small to permit a meaningful determination of OPR$_{678}$ and $T_{86}.$  Given the limitations of our simple model, the agreement is quite good.  Thus the changing OPR, witnessed here for the first time, is broadly consistent with the predictions of interstellar shocks models.  { In particular, the data provide among the most compelling evidence yet obtained for C-type shocks in which 
the flow velocity changes continuously.}

\begin{acknowledgements}

Based on observations made with the NASA/DLR Stratospheric Observatory for Infrared Astronomy (SOFIA). SOFIA is jointly operated by the Universities Space Research Association, Inc. (USRA), under NASA contract NAS2-97001, and the Deutsches SOFIA Institut (DSI) under DLR contract 50 OK 0901 to the University of Stuttgart. D.A.N gratefully acknowledges the support of an USRA grant, SOF06-0022;  
and S.C., A.G., P.L. and L.N.T. that of the Programme National ``Physique et
Chimie du Milieu Interstellaire" (PCMI) of CNRS/INSU with INC/INP co-funded by CEA and CNES.

\end{acknowledgements}

\end{document}